\documentclass[aps,superscriptaddress,,twocolumn,showpacs]{revtex4-1}

\pdfoutput=1

\usepackage{color}
\usepackage{amsmath, amsthm, amsfonts}    
\usepackage{amssymb}
\usepackage{mathptmx} 
\usepackage{dsfont}
\usepackage{tikz}

\usepackage[unicode=true,pdfusetitle,
 bookmarks=true,bookmarksnumbered=false,bookmarksopen=false,
 breaklinks=true,pdfborder={0 0 0},backref=false,colorlinks=true,citecolor=blue]{hyperref}

\usepackage{graphicx}   
\usepackage[caption=false]{subfig}       
\usepackage{bm}            
\usepackage[normalem]{ulem}  

\newlength\imagewidth
\newlength\imagescale

\def\be{\begin{eqnarray}}
\def\ee{\end{eqnarray}}
\def\r{{\bf r}}

\def\E{{\bf E}}
\def\H{{\bf H}}

\def\im{{\rm i}}

\def\tensorS{\overset{\text{\scriptsize$\leftrightarrow$}}{{\bf S}}}

\definecolor{JOT-color}{named}{blue}
\definecolor{CSF-color}{named}{orange}

\begin{document}

\title{Multiple Kerker conditions in  arbitrary dielectric spheres}

\author{Cristina Sanz-Fern\'andez}
\email{cristina$_$sanz001@ehu.eus}
\affiliation{Centro de F\'{i}sica de Materiales (CFM-MPC), Centro Mixto CSIC-UPV/EHU,  20018 Donostia-San Sebasti\'{a}n, Spain}

\author{Jorge Olmos-Trigo}
\email{jolmostrigo@gmail.com}
\affiliation{Donostia International Physics Center (DIPC),  20018 Donostia-San Sebasti\'{a}n, Spain}

\author{Juan Jos\'e S\'aenz}
\affiliation{Donostia International Physics Center (DIPC),  20018 Donostia-San Sebasti\'{a}n, Spain}
\affiliation{IKERBASQUE, Basque Foundation for Science, 48013 Bilbao, Spain}

\begin{abstract}
High refractive index dielectric particles present unique light scattering properties in the spectral range dominated by electric and magnetic dipolar resonances. These properties are absent in non-resonant low-index particles  due to the strong overlap between dipolar and  higher order multipolar contributions. Here we propose to induce duality and zero optical backscattering (first Kerker conditions) at multiple size parameter values on arbitrary dielectric spheres by illuminating with a tightly focused dipolar beam. As an illustrative example, we show that duality signatures, including topological optical vortices and enhanced spin orbit optical mirages,
can be  induced on arbitrary sized spheres of Polystyrene immersed in water.
\end{abstract}

\maketitle

High refractive index (HRI) subwavelength particles  have received an increasing interest as building blocks of new all-dielectric metamaterials and optical devices~\citep{kuznetsov2016optically}. Their remarkable optical properties are associated to  the simultaneous excitation  of well-defined electric and magnetic dipolar resonant modes~\citep{evlyukhin2010optical,garcia2011strong,gomez2012electric,Shi2013}. These include  strong backward-to-forward asymmetric scattering and, at the first Kerker condition~\citep{kerker1983electromagnetic,nieto2011angle}, {when the duality symmetry is satisfied~\citep{zambrana2013duality}}, the emergence of  zero optical backscattering~\citep{gomez2011electric,geffrin2012magnetic,fu2013directional,person2013demonstration}.  The scattering directionality  can be tuned further by coupling dipolar electric and magnetic modes with quadrupolar responses~\citep{alaee2015generalized,shamkhi2019transverse}. 
{Under circularly polarized illumination, HRI nanoparticles can present important spin-orbit interactions (SOI)~\citep{bliokh2015spin}.
At the first Kerker condition, these interactions lead to  a backscattering topological vortex~\citep{olmos2019enhanced}, which in turn reveals, due to the spiraling structure of the scattered light, a diverging apparent  displacement in the position of the particle (SOI optical mirages)~\citep{olmos2019enhanced,Olmos2018SOI,gao2018enhanced}}. 
Our main goal  is to show that, under appropriate illumination, all these interesting scattering phenomena, usually associated to resonant HRI particles,  can be observed in arbitrary dielectric particles, regardless  their  size  and  refactive index contrast.

Most of previous works implicitly assume plane wave (PW) illumination conditions, although, as well as the polarization, the spatial structure of the incoming light field plays a major role in the excitation of the nanoparticle resonances~\citep{zumofen2008perfect,mojarad2008plasmon,zambrana2012excitation,wozniak2015selective,
zambrana2018tailoring,bag2018transverse,eismann2018exciting,nechayev2019orbital, zaza2019size}. 
On the other hand, illumination with a tightly focused beam can still only excite particle multipolar modes which are already present in the incident beam.
Tailoring the multipolar expansion of the focused beam would then enable a detailed study of selected individual multipolar resonances without interference by the other modes. For instance, cylindrical beams carrying a  well-defined angular momentum (AM) component along the beam axis~\citep{allen1992orbital}, $m\hbar$, can only be coupled with multipolar modes with total AM number $l \ge |m|$ (where $l=1$ correspond to dipolar resonances, $l=2$ to quadrupolar, etc.). These beams have then been used to suppress the contribution of  low order multipolar resonances~\citep{zambrana2012excitation} enabling the excitation of Kerker conditions (dual modes)  of some  high order multipoles. 
In this vein, by tuning the field patterns and polarization of tightly focused beams it is possible to selectively  excite  either electric or magnetic low order modes of HRI Silicon particles~\citep{wozniak2015selective} or a combination of  spectrally overlapping dipolar resonances leading to the  transverse Kerker scattering~\citep{bag2018transverse}.
However, in contrast with HRI particles, achieving dipolar Kerker conditions for ubiquitous low refractive index (LRI) dielectric particles,  (e.g. standard Polystyrene -PS-  spheres)  is a challenging problem~\citep{abdelrahman2019experimental}. The strong spectral overlap between the dipolar electric and magnetic modes, and a broad range of  higher order multipoles, makes these conditions to remain hidden \footnote{Interestingly, dipolar resonances of low-index spheres can be enhanced by  partial metallic dressing~\citep{zhou2018visualizing}.}.

In this Letter, we show that multiple Kerker conditions can be induced in arbitrary dielectric spheres, regardless their size  and refractive index contrast, by illuminating with a dipolar tightly focused  beam with a well-defined total AM number, $l = 1$. In analogy with HRI spheres, the scattering can  be tuned from almost zero forward radiated power to perfect zero-backscattering at multiple first Kerker conditions, where the particle’s response is dual. For dipolar beams with well-defined helicity~\citep{olmos2019sectoral}, it is possible to induce dual scattering  signatures, including topological optical vortices and enhanced optical mirages on arbitrary dielectric spheres.
As a specific example, we discuss in detail multiple dipolar Kerker conditions from the scattering of  Polystyrene beads immersed in water.

\begin{figure}
\includegraphics[width=1 \columnwidth]{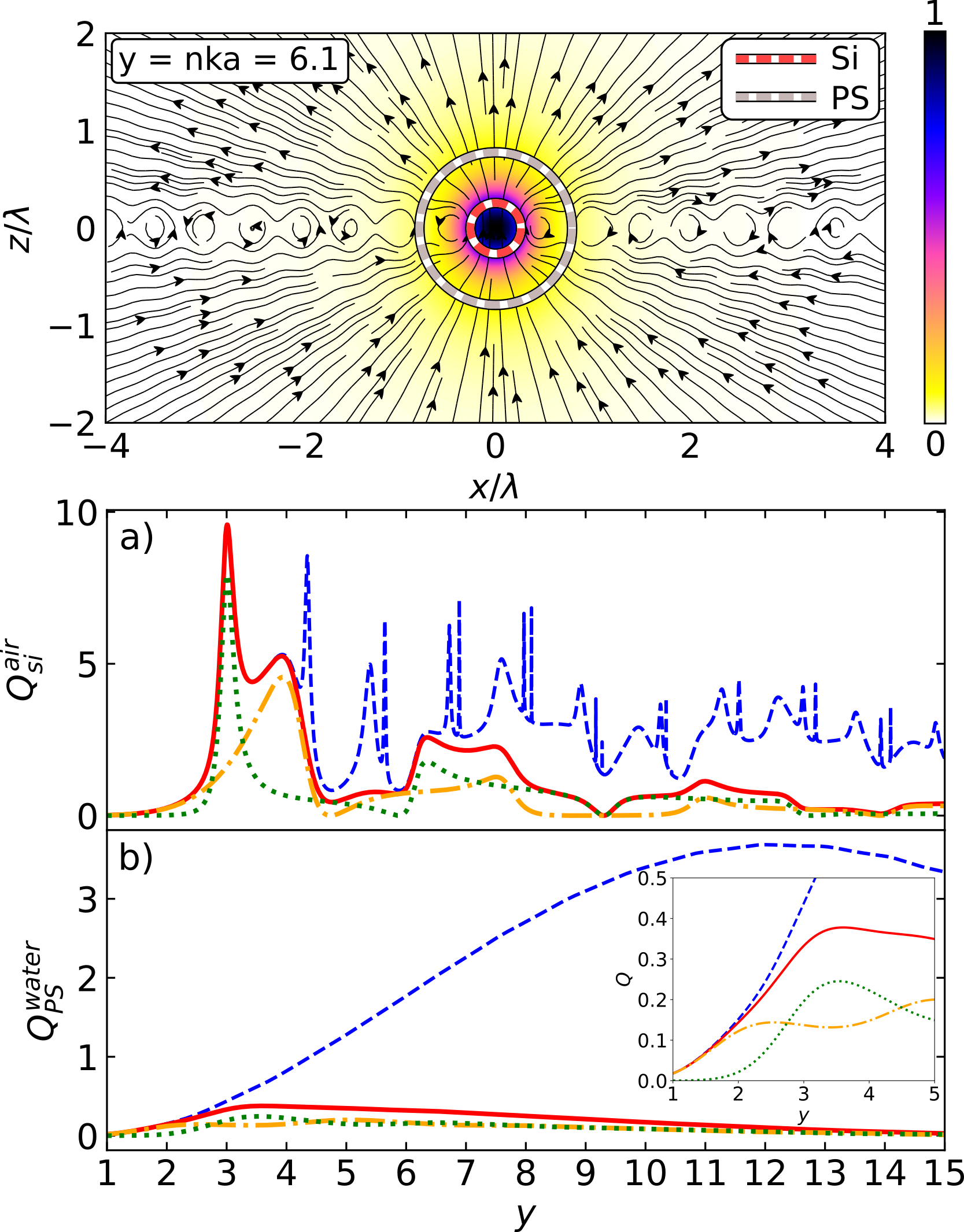}
\captionsetup{justification= raggedright}
\caption{Projected  flow lines of the Poynting vector for  ($\boldsymbol{\Psi}_{11}^{+1}$) pure dipole beam. The intensity color map corresponds to the field intensity normalised to the maximum field intensity at the origin (after Ref.~\citep{olmos2019sectoral}). Dashed circles correspond to  Silicon-like (red) and PS (grey) spheres for $y=6.1$. Scattering efficiency, $Q$ versus the $y$-size  parameter, $y = n 2\pi a/\lambda $, (dashed -blue- lines) for a dielectric sphere of radius $a$ made of  (a) Silicon-like spheres  in air ( $n=  3.46$) and (b) Polystyrene  immersed in  water
($n = 1.2$). 
Dotted (green) and dashed-dotted (orange) lines correspond to the dipolar magnetic and dipolar electric contributions respectively, while their sum, 
solid (red) lines, are the total dipolar contribution to the scattering efficiency. The inset in (b) is  a zoom  of $Q_{\text{PS}}$  showing the relative weight of electric and magnetic dipolar modes  for relatively low values of the $y$-parameter.  \label{cross}}
\end{figure}

Let us first consider an arbitrary  monochromatic light beam 
(with an implicit time-varying harmonic component $e ^{-iwt}$) propagating through an homogeneous  medium with real refractive index $n_{\rm{h}}$ and wave number $k=n_{\rm{h}} k_0 = 2 \pi n_{\rm{h}} / \lambda_0$ (being $\lambda_0$ the light wavelength in vacuum).
Within a helicity and  AM framework~\citep{fernandez2012helicity}, the beam electric field, $\E_{\text{inc}}$, can be expanded in vector spherical wavefunctions (VSWFs), $\boldsymbol{\Psi}_{lm}^{\sigma}$,  with well-defined helicity,  $\sigma = \pm 1$~\citep{olmos2019enhanced}. In this basis, 
\be
\E_{\text{inc}} &=& \sum_{\sigma= \pm 1} \E_{\text{inc}}^\sigma, \quad  \E_{\text{inc}}^\sigma = E_0 \sum_{l=1}^{\infty} \sum_{m=-l}^{+l}  C_{lm}^{ \sigma} \boldsymbol{\Psi}_{lm}^{\sigma},
\label{multipolar} 
\ee
where $\boldsymbol{\Psi}_{lm}^{\sigma}$ is defined as
\be
\boldsymbol{\Psi}_{lm}^{\sigma} &=& \frac{1}{\sqrt{2}} \left[ {\boldsymbol{N}}_{lm} +  \sigma {\boldsymbol{M}}_{lm}  \right], \label{V1} \\
{\boldsymbol{M}}_{lm} &\equiv & j_l(kr)\boldsymbol{X}_{lm}  \quad, \quad
{\boldsymbol{N}}_{lm} \equiv \frac{1}{k} \boldsymbol{\nabla} \times {\boldsymbol{M}}_{lm}, \label{V3} \\
\boldsymbol{X}_{lm} &\equiv& \frac{1}{\sqrt{l(l+1)}} {\bf{L}} Y_l^m (\theta,\varphi), \label{V3}
\ee
being $\boldsymbol{M}_{lm}$ and $\boldsymbol{N}_{lm}$  are Hansen's multipoles~\citep{borghese2007scattering}, $\boldsymbol{X}_{lm}$ denotes the vector spherical harmonic~\citep{jackson1999electrodynamics}, $ j_l(kr)$ are the spherical (well-defined at $r = 0$) Bessel functions,  $Y_l^m$ are the spherical harmonics, and $ {\bf{L}} \equiv  \left\{ -\im \r \times \boldsymbol{\nabla}\right\} $ is the orbital angular momentum (OAM) operator. 
Let us recall that the multipoles $ \boldsymbol{\Psi}_{lm}^{\sigma} $ 
can be built following the standard rules of angular momentum addition~\citep{borghese2007scattering,edmonds2016angular}
as simultaneous eigenvectors of the square of the total angular momentum,   $\bf{J}^2$, and its $z$-component,  $J_z$, with $\bf{J} = {\bf{L}} + \tensorS$, being $\tensorS \equiv -\im {\bf{I}} \times$ the  spin angular momentum (SAM) operator and ${\bf{I}}$ the unit dyad. The multipoles $ \boldsymbol{\Psi}_{lm}^{\sigma} $ are then simultaneous eigenvectors of $\bf{J}^2$, $J_z$~\citep{edmonds2016angular} and the helicity operator~\citep{fernandez2012helicity} $\boldsymbol{\Lambda} = (1/k) \boldsymbol{\nabla} \times$, with eigenvalues $l(l+1)$, $m$ and $\sigma$, respectively.

When  a homogeneous dielectric sphere of radius $a$  and an arbitrary (absorptionless) refractive index $n_{\rm{p}}$ is  centred at the origin ($r = 0$), the AM numbers, $l$ and $m$, are preserved in the scattering process due to the symmetry of the system.  The scattered fields outside the sphere, $\E_{\text{sca}} = \E_{\text{sca}}^+ + \E_{\text{sca}}^-$, can be written in terms of ``outgoing'' VSWFs, $\bm{\Phi}_{lm}^{\sigma'}$ (defined as in Eq.~\eqref{V1} replacing $j_l(kr)$ by the outgoing spherical Bessel functions $h_l(kr)$), as
\be
\E_{\text{sca}}^\sigma &=& E_0 \sum_{l=1}^{\infty} \sum_{m=-l}^{+l}  D_{lm}^{ \sigma} \boldsymbol{\Phi}_{lm}^{\sigma}, \nonumber \\
\begin{pmatrix} D_{lm}^+ \\ D_{lm}^- \end{pmatrix} &=&  
-\begin{pmatrix} [a_l+b_l] & -[a_l-b_l] \\ [a_l-b_l] & -[a_l +b_l]  \end{pmatrix} 
\begin{pmatrix} C_{lm}^+ \\   C_{lm}^- \end{pmatrix}, \label{helimatrix}
\label{multipolar_scatter} 
\ee
where $Z \H_{\text{sca}}^\sigma= -\im \sigma  \E_{\text{sca}}^\sigma$ with $Z=1/(\epsilon_0 c n_{\rm{h}})$, and $a_l$ and $b_l$ are the electric and magnetic Mie coefficients~\citep{bohren2008absorption}. 

In this framework, it can be seen that the expected value of the helicity of the scattered field is generally given by
\be \label{V:5} 
\langle \Lambda \rangle &=& 
\frac{ \langle \bm{E}_{\rm{sca}}^* \cdot \left(  \bm{\Lambda} \bm{E}_{\rm{sca}} \right) \rangle } 
       {\langle \bm{E}_{\rm{sca}}^* \cdot \bm{E}_{\rm{sca}} \rangle }
=  \frac{\sum_{l=1}^\infty \sum_{m=-l}^{+l} |D_{lm}^+|^2 - |D_{lm}^-|^2  }{\sum_{l=1}^\infty \sum_{m=-l}^{+l} |D_{lm}^+|^2 + |D_{lm}^-|^2 }.
\ee 
When the sphere is illuminated by a circularly polarized PW with helicity $\sigma$,
or by a cylindrically symmetric beam (eigenvector of  the helicity, $\bm{\Lambda}$  and $z$-component of the AM, $J_z$,  operators  with eigenvalues  $m$ and $\sigma$ ),
the scattered field is, in general, a combination of multipolar modes with fixed $m$. As a result, it can be shown that, 
\begin{align} \label{V:5} 
\langle \Lambda \rangle = \sigma \frac{1-T}{1+T} , &&\text{where} &&  T = \frac{\sum_{l=|m|}^\infty    \left| C_{lm}^+  \right|^2   |a_l -b_l|^2}{\sum_{l=|m|}^\infty  \left| C_{lm}^+  \right|^2 |a_l + b_l|^2 } 
\end{align}
is the helicity transfer function~\citep{zambrana2013dual}.
When $T$ goes to zero, the particle response is dual and the scattered helicity is the same as the helicity of the incoming beam. 
However, the helicity is not preserved in a \emph{multipolar scattering process} except for the  case $a_l=b_l, \ \forall \ l$, which can only be achieved for hypothetical spheres with equal electric permittivity $\epsilon$ and magnetic permeability $\mu$, corresponding to the original first Kerker condition~\citep{kerker1983electromagnetic}. 

To address this issue, we propose the use of dipolar beams (DB), that present a well-defined total AM, i.e., $l = 1$. Under this illumination, which is solution of Maxwell’s equations, the multipolar response
is by construction discarded, regardless of the $x = ka$ size
parameter and the relative refractive index $n = n_{\rm{p}} /n_{\rm{h}}$. 
The most general DB can be written as a combination of pure dipole beams (PDB)  with well-defined AM  and helicity eigenvalues $(l=1, m, \sigma=\pm 1)$~\citep{olmos2019sectoral},
$\E_{\text{inc}}^D = \E_{\text{inc}}^{D+} + \E_{\text{inc}}^{D-}$, where
 \be
\E_{\text{inc}}^{D\sigma}(\r) = E_0 \left(C_{11}^\sigma \boldsymbol{\Psi}_{11}^{\sigma}+C_{10}^\sigma \boldsymbol{\Psi}_{10}^{\sigma}+C_{1-1}^\sigma \boldsymbol{\Psi}_{1-1}^{\sigma}\right) .
\ee
At the focus $r=0$, it takes the simple form
\be
\E_{\text{inc}}^{D\sigma}(0)  = \im \frac{ E_0}{\sqrt{12 \pi}} \left(C_{11}^\sigma \bm{\xi}_1+C_{10}^\sigma \bm{\xi}_0 +C_{1-1}^\sigma \bm{\xi}_{-1} \right),
\ee
with
$
\bm{\xi}_1 = - ( \bm{\hat{x}} + \im \bm{\hat{y}}) /\sqrt{2}, \quad \bm{\xi}_0= \bm{\hat{z}}, \quad \bm{\xi}_{-1} =  ( \bm{\hat{x}} - \im \bm{\hat{y}})/\sqrt{2}$. When a dielectric sphere is illuminated by a PDB mode, as e.g. (see top panel in Fig.~\ref{cross})
 \be
\E_{\text{inc}}^D = E_0 \im \sqrt{12 \pi}  \boldsymbol{\Psi}_{1\sigma}^{\sigma}, \label{pdbs} \ee
or by a linear combination of opposite helicities
 \be
\E_{\text{inc}}^D = E_0 \im \sqrt{6 \pi} \left( \boldsymbol{\Psi}_{11}^{1} - \boldsymbol{\Psi}_{1-1}^{-1}\right), \label{dbl} \ee
where 
\be
C_{lm}^{\sigma} =  \im^l \sqrt{4 \pi(2l+1)} \delta_{m\sigma},
\ee
the electric and magnetic fields at the focus are transversal with respect to the optical $z$-axis. Therefore, at the focus, these fields behave 
exactly the same as those induced by a circularly polarized PW of helicity $\sigma$, or by a linearly polarized PW, respectively.
As a direct consequence, for small dipolar particles located at the focus, 
the dipolar scattered fields are strictly equivalent. 
Other combinations of PDBs, including  $\boldsymbol{\Psi}_{11}^{\sigma}$ and $\boldsymbol{\Psi}_{10}^{\sigma}$ modes,
can be used to induce focal fields with both longitudinal and transversal components leading to the so-called transversal Kerker conditions~\citep{bag2018transverse}. In this case, the longitudinal fields arise from the interference between $\boldsymbol{\Psi}_{11}^{\sigma}$ and $\boldsymbol{\Psi}_{10}^{\sigma}$ modes since $\boldsymbol{\Psi}_{10}^{\sigma}$ modes do not carry power by themselves~\citep{olmos2019sectoral}. For the same reason, it is not possible to induce pure longitudinal electric and magnetic fields with a dipolar beam. 
According to the aforementioned, the scattering that arises from a PDB (or a linear combination of them) can be easily understood by comparison with the scattering that emerges from PW illumination.

\begin{figure}
\includegraphics[width=1 \columnwidth]{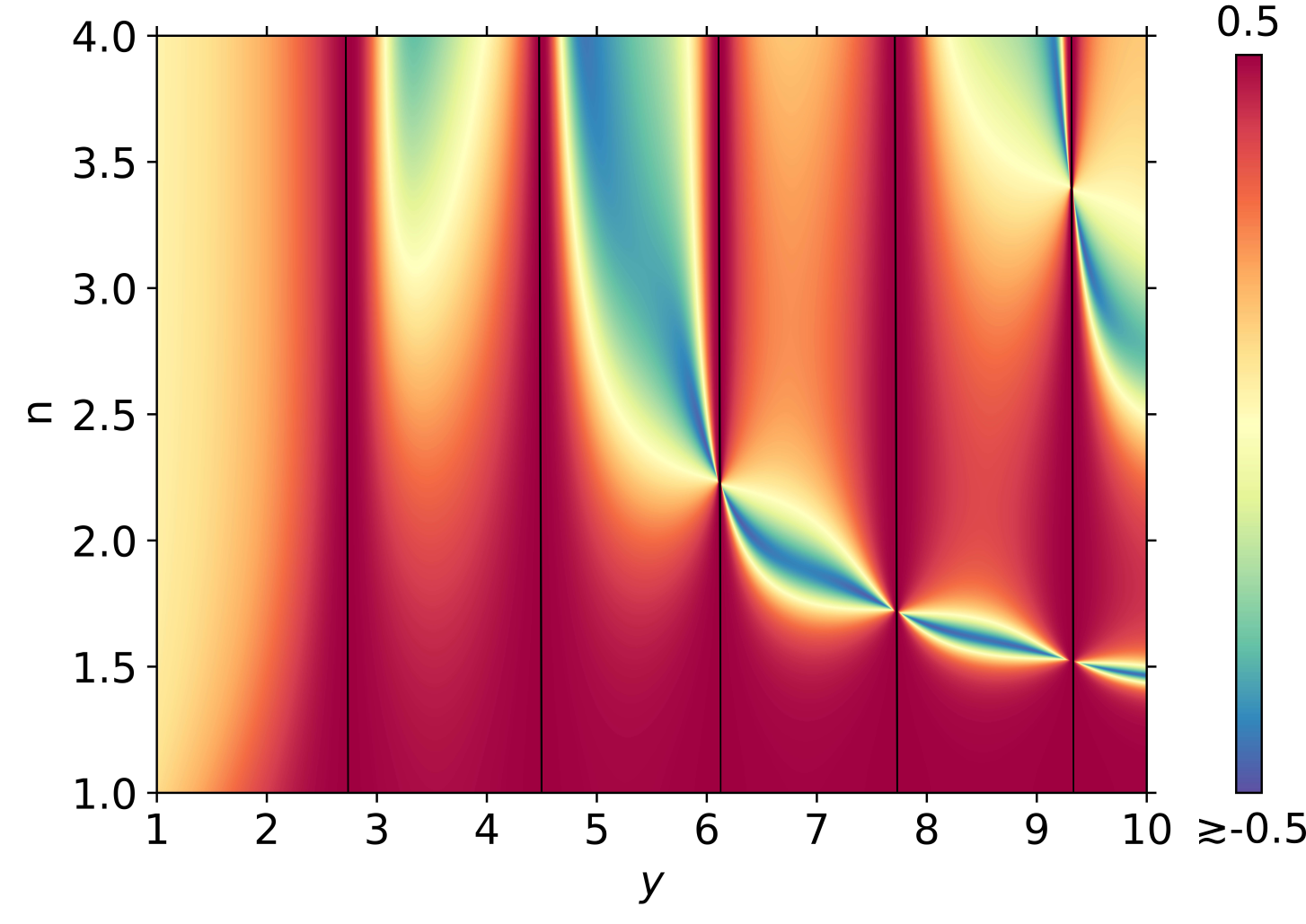}
\captionsetup{justification= raggedright}
\caption{Color map of the asymmetry parameter, $g$, when a dielectric sphere is excited with a  dipolar beam  (see Eqs.~\eqref{pdbs} or \eqref{dbl} ) as a function of the relative refractive index $n$ and size parameter $y = nka$.  The scattering becomes dual at successive  Kerker conditions $a_1=b_1$  (black vertical lines) where $g$ takes its maximum value of $g=1/2$ and the scattering is dual. The minimum (analytical) value, given by $g = -1/2$, cannot be reached  due to causality~\citep{alu2010does}.  }\label{1}
\end{figure}

Figure~\ref{cross} (bottom panel) illustrates the behaviour of the scattering efficiency $Q$~\citep{bohren2008absorption}, 
versus the size parameter $y= n k a $, when typical examples of HRI and LRI lossless spheres are illuminated with  both a PW (dashed blue) and a PDB (solid red line), Eqs.~\eqref{pdbs} and~\eqref{dbl}. $Q \equiv \sigma_{\rm{sca}} / (\pi a^2)$ is proportional to the scattering cross section, $\sigma_{\rm{sca}}$, defined as the ratio between the total scattered power and the incident energy flux.
Figure~\ref{cross} (a) corresponds to $n = \sqrt{12} $ (similar to Si-like spheres in air -see~\citep{fenollosa2008silicon}-) and Fig.~\ref{cross} (b) to $n =  1.56/1.3 =1.2$ (PS in water). As it can be inferred from the comparison between the PW and PDB responses, HRI Si-like spheres  present a  clear dipolar response for size parameters $y= n k a \lesssim 4.5$. It is in this spectral region where it is possible to observe the dipolar Kerker condition $a_1=b_1$. For larger sizes, the overlapping with higher order multipoles masks the  sequential crossing between electric and magnetic dipolar responses (dotted green and dash-dotted yellow lines), which would lead to dipolar Kerker conditions at multiple $y$ size parameters~\citep{zambrana2013dual}. In contrast, Fig.~\ref{1} (b) illustrates that, for non-resonant LRI spheres, dipolar dual conditions cannot be observed  due  the strong overlap between the  dipolar and higher order multipoles even at low $y$ sizes.
  
We shall now move to the main consequences of this DB illumination, based on the unmasking of the hidden multiple dipole Kerker conditions. To that end, it results convenient to introduce the asymmetry parameter~\citep{gomez2012negative,pinheiro2000new}, defined as the average of the cosine over the scattering angle, 
\be
g= \langle \cos \theta \rangle = \frac{\text{Re}\left\{a_1b_1^*\right\} } {  |a_1|^2+|b_1|^2}, \quad -1/2 < g \le 1/2, \label{g}
\ee
which encodes the interference between the first electric and magnetic Mie coefficients. 
The maximum value, $g=1/2$, corresponds to the Kerker zero-backscattering condition $a_1=b_1$. 
This condition, well-known for small dipolar particles under PW illumination, can now be extended to arbitrary sized dielectric spheres illuminated by a (sectoral) dipolar beam \citep{olmos2019sectoral}, regardless of the incoming helicity. A map of  $g$ for a dielectric sphere as a function of the refractive index contrast $n$ and the $y$ size parameter, see Fig.~\ref{1}, reveals the existence of successive first  Kerker conditions  even for LRI dielectrics. This phenomenon goes well beyond of previous works which were based on Kerker conditions in the limit of small particle~\citep{gao2017optical}.

\begin{figure}[]
\includegraphics[width = 1\columnwidth]{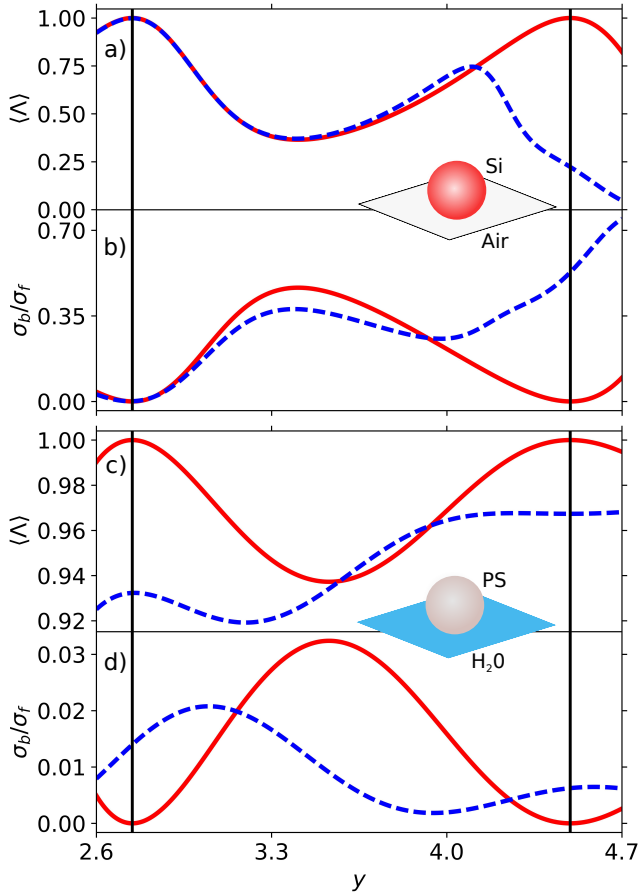}
\captionsetup{justification= raggedright}
\caption{
(a) Expected value of the helicity of the scattered field, $\langle \Lambda \rangle$,  versus the $y$-size parameter and (b)
the corresponding ratio between backscattering and forward scattering differential cross sections, $\sigma_b/\sigma_f$, for a Silicon-like sphere in air under 
PW (dashed -blue- line) and pure dipolar beam ( PDB, continuous -red- line ) illumination with helicity $\sigma=+1$.
(c) and (d) show the same quantities for a PS sphere immersed in air. 
Dipolar illumination induces a dual response at 
$y \approx 2.7$ and $y \approx 4.5$ (Vertical -black- lines) 
at which the helicity is preserved ($\langle \Lambda \rangle=\sigma=+1$) and the backscattering intensity is zero.
In contrast,  PW only induces duality for HRI at $y \approx 2.7$ where the scattering is dominated by the electromagnetic dipolar response.} \label{2}
\end{figure} 

\begin{figure}[t]
\includegraphics[width=1 \columnwidth]{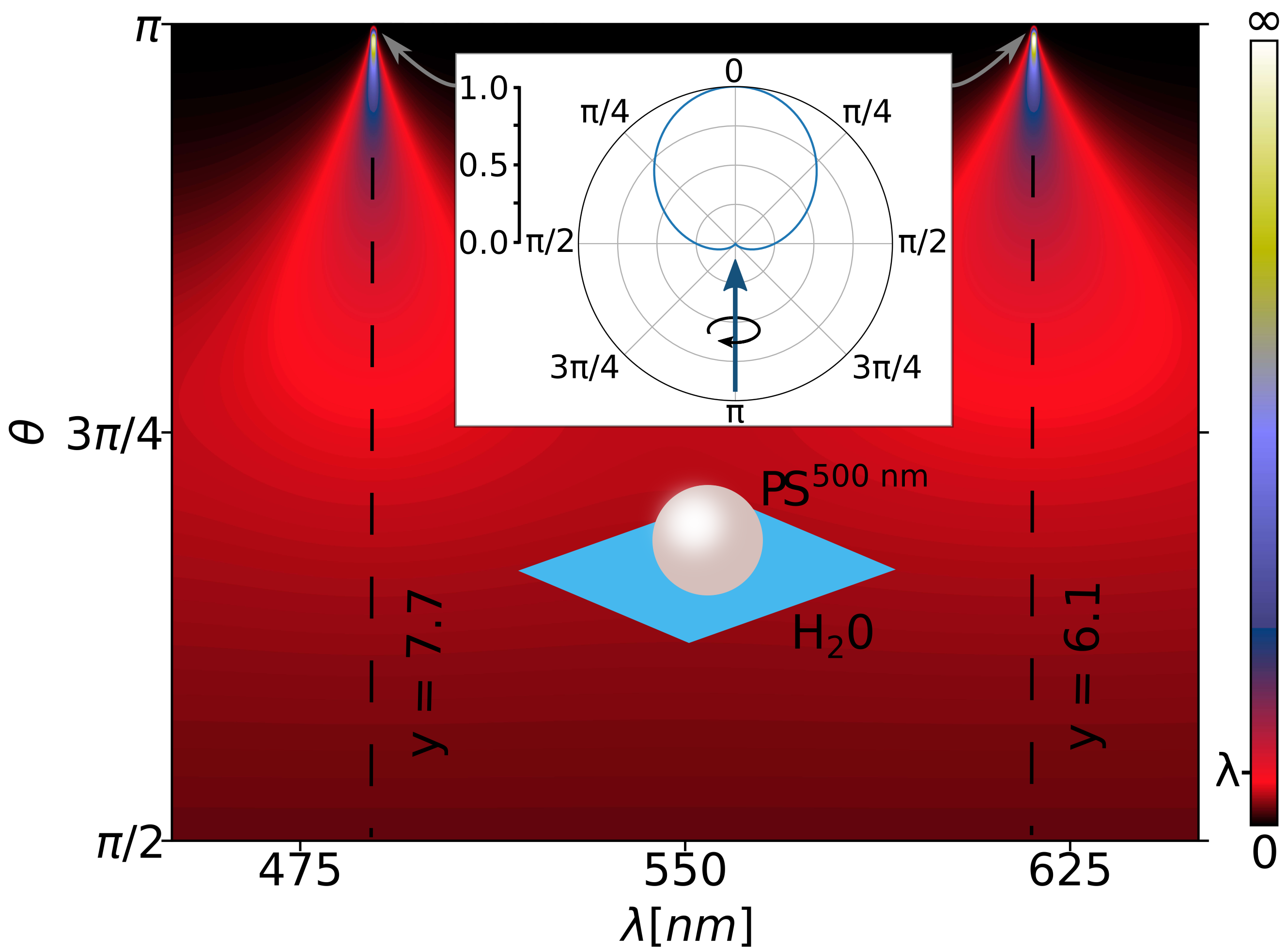}
\captionsetup{justification= raggedright}
\caption{Optical mirage ($\Delta / \lambda$) originated from the scattering of a 500 nm PS sphere immersed in water when exciting with a PDB  in the visible spectral range. The wavelength $\lambda = \lambda_0 /  n_{\rm{h}}$, where $n_{\rm{h}} = 1.33$ (water). Diverging values are reached at backscattering, $\theta = \pi$,  at the first oscillating Kerker condition. The inset shows the normalized differential cross section divided by the total cross section when illuminating with a PDB. 
}\label{3}
\end{figure}

Under illumination by a PDB of well-defined helicity, the remarkable angular dependence of the intensity and the averaged helicity are linked, and fully determined by the dipolar asymmetry parameter $g$, Eq.~\eqref{g}~\citep{olmos2019asymmetry}. Particularly, the helicity transfer function $T$ is now given by the dipolar term in Eq.~\eqref{V:5}, $l =  m = \sigma$, coinciding with the ratio between backward and forward cross sections,
\be T = \frac{ \sigma_b}{\sigma_f} = \frac{|a_1-b_1|^2}{|a_1+b_1|^2} = \frac{1- \sigma \langle \Lambda \rangle}{1+ \sigma \langle \Lambda \rangle} = \frac{1- 2g}{1+ 2g} \label{bf},
\ee
where we have used the dipolar results for the expected value of the helicity, $\langle \Lambda \rangle = 2\sigma g$~\citep{olmos2019asymmetry}.

Figure \eqref{2} illustrates the behaviour of both $\langle \Lambda \rangle$ and $T= \sigma_b/\sigma_f$ for a Si-like sphere in air and a PS sphere immersed in water, under PW and PDB illumination with well-defined $\sigma = +1$. Through PW illumination, the duality symmetry ($\langle \Lambda \rangle=+1$), which traduces into zero optical backscattering ($ \sigma_b/\sigma_f=0$), can only be achieved in the limit of small particle for the Si case, according to Fig.~\eqref{3} (a) and (b), respectively. However, under PDB, the unmasked first Kerker conditions ($g=1/2$) shown in Fig.~\eqref{2} induce successively these phenomena for both examples. Notice that in the latter case the first Kerker condition is perfectly achieved, whereas under PW illumination, a slight contribution of higher multipoles may always induce a (despicable) suppression of the duality symmetry.

Interestingly, recent predictions concerning SOI phenomena in the dipolar approximation~\citep{olmos2019enhanced}, limited to subwavelength HRI particles, can also be extended to arbitrary spheres by PDB illumination. For a PDB with well-defined helicity, e.g. $\sigma = +1$, the (dimensionless) $z$-component of the AM, $m=\sigma$, can be written as the sum of the OAM, $l_z(\theta)$, and SAM, $s_z(\theta)$, densities $m=l_z+s_z$, which is conserved in the scattering process for axially symmetric scatterers~\citep{moe1977conservation,schwartz2006conservation,bliokh2011spin}. 
In this case the OAM density of the scattered photons is given by~\citep{olmos2019enhanced}
\be
l_z(\theta) = \frac{ \sin^2 \theta \left(1+ 2g \cos \theta \right)}{1 + \cos^2 \theta +4 g\cos \theta},
\ee
which approaches $l_z=+2$, i.e., $s_z=-1$, near backscattering, signaling the emergence of a topological vortex at every Kerker condition ($g=1/2$). The spiraling structure of the scattered light, associated to $l_z(\theta)$, is responsible of the apparent  displacement in the position of the particle (SOI optical mirages)~\citep{olmos2019enhanced,Olmos2018SOI,gao2018enhanced} given by
$\Delta/ (\lambda/ \pi) = l_z/\sin \theta$. 
Figure~\ref{3} illustrates the optical mirage induced from a PDB  impinging on a $0.5\ \mu m$ PS sphere immersed in water in the visible spectral range. As it can be seen, the SOI optical mirage is considerably enhanced near the backscattering angles ($\theta \sim \pi$) at the successive first Kerker conditions, where the zero-backscattering is fulfilled as shown in the figure inset.

In conclusion, we have shown that dipolar beams with well-defined helicity can induce multiple Kerker conditions and duality symmetry to  arbitrary  dielectric spheres. Duality signatures and spin-orbit effects like perfect zero optical backscattering, emergence of  topological vortex and enhanced optical mirages, predicted for subwavelength HRI spheres, could be experimentally observed in LRI particles like standard, micron-sized, Polystyrene spheres immersed in water.

This research was supported by the Spanish Ministerio
de Econom\'{\i}a y Competitividad (MINECO, MICINN)
and European Regional Development Fund (ERDF)
Projects FIS2015-69295-C3-3-P and FIS2017-82804-P, and by the Basque Dep. de Educaci\'on
Project PI-2016-1-0041 and PhD Fellowship (PRE-
2018-2-0252).

\bibliography{New_era_18_06_2019}
\end{document}